\documentclass[12pt]{article}
\usepackage[utf8]{inputenc}

\usepackage{comment}
\usepackage{float} 
\usepackage{bm}
\usepackage{natbib}
\usepackage{subcaption}
\usepackage{setspace}
\usepackage[ruled,linesnumbered]{algorithm2e}
\usepackage{xcolor}
\usepackage{graphicx,amsfonts,amsthm,amssymb}
\usepackage[colorlinks,citecolor=blue,urlcolor=blue]{hyperref}
\usepackage{amsmath,mathrsfs}
\usepackage{wrapfig,lipsum,booktabs}
\usepackage{appendix}
\usepackage{enumitem}
\usepackage{multirow}
\usepackage{comment}
\usepackage{hyperref}
\usepackage[font=footnotesize,skip=0pt]{caption}
\usepackage{amssymb}
\usepackage{pifont}
\newcommand{\cmark}{\ding{51}}%
\newcommand{\xmark}{\ding{55}}%

\newcommand{\blind}{1}

\def\spacingset#1{\renewcommand{\baselinestretch}%
{#1}\small\normalsize} \spacingset{1}

\def\Pr{{\mathbb{P}}}
\def\Ex{{\mathbb{E}}}

\def\bxi{\boldsymbol\xi}

\def\0{{\bf 0}}
\def\1{{\bf 1}}

\def\D{{\bf D}}

\def\pr{ \mathbb{P}}

\def\0{{\bf 0}}

\def\dfrac#1#2{{\displaystyle{#1\over#2}}}

\def\bq{\begin{equation}}
\def\eq{\end{equation}}

\def\log{\mathrm{log}}

\def\squarebox#1{\hbox to #1{\hfill\vbox to #1{\vfill}}}

\def\bx{{\bf x}}

\def\dfrac#1#2{{\displaystyle{#1\over#2}}}

\def\bse{\begin{eqnarray*}}
	\def\ese{\end{eqnarray*}}
\def\be{\begin{eqnarray}}
	\def\ee{\end{eqnarray}}
\def\bsq{\begin{equation*}}
	\def\esq{\end{equation*}}
\def\bq{\begin{equation}}
	\def\eq{\end{equation}}

\def\boxit#1{\vbox{\hrule\hbox{\vrule\kern6pt\vbox{\kern6pt#1\kern6pt}\kern6pt\vrule}\hrule}}

\def\frakN{{\mathfrak{N}}}

\usepackage[bottom]{footmisc}
\newtheorem{theorem}{Theorem}
\newtheorem{example}{Example}

\def\pr{\mathbb{P}}
\def\Ex{\mathbb{E}}

\renewcommand{\appendix}{
 \setcounter{section}{0}%
  \setcounter{subsection}{0}%
  \renewcommand\thesection{\Alph{section}}
  \setcounter{equation}{0}
  \renewcommand{\theequation}{S.\arabic{equation}}
  \setcounter{figure}{0}
  \renewcommand\thefigure{S\arabic{figure}}
  \setcounter{table}{0}
  \renewcommand\thetable{S\arabic{table}}  
  }

\allowdisplaybreaks
\begin{document}

\if1\blind
{
  \title{
 Distribution-free screening of spatially variable genes in spatial transcriptomics
  }	
 \author{Changhu Wang \thanks{Equal contribution}  \\
     Qiyun Huang\footnotemark[1] 
    \\ Zihao Chen \footnotemark[1] \\
    School of Mathematical Sciences, Peking University\\
    Jin Liu \thanks{Co-corresponding author}\\
 School of Data Science\\
 The Chinese University of Hong Kong, Shenzhen\\
    Ruibin Xi \footnotemark[2]\\
      School of Mathematical Sciences\\
      Center for Statistical Science, 
      Peking University
}
  \date{}
  \maketitle
} \fi

\if0\blind
{
  \title{
 Distribution-free screening of spatially variable genes in spatial transcriptomics
  }	
    \date{}
  \maketitle

  \medskip
} \fi

\maketitle

\def\spacingset#1{\renewcommand{\baselinestretch}%
{#1}\small\normalsize} \spacingset{1}


\begin{abstract} 
Spatial transcriptomics (ST) technologies enable transcriptome-wide gene expression profiling while preserving spatial resolution, offering unprecedented opportunities to uncover complex spatial structures. Due to the ultra-high dimensionality of ST data, identifying spatially variable genes (SVGs) associated with unknown spatial clusters has become a central task in ST data analysis. Here, we develop a distribution-free SVG screening method based on a novel quasi-likelihood ratio statistic, the MM-test, combined with a knockoff procedure to control the false discovery rate (FDR). MM-test leverages auxiliary information, such as spatial distances, about the unknown spatial domains for SVG screening. Notably, in addition to two-dimensional ST datasets, MM-test is well-suited for increasingly common three-dimensional (3D), multi-slice ST datasets. Extensive benchmarking using simulations and 34 real ST datasets demonstrates that MM-test consistently outperforms existing SVG detection methods. In a 3D mouse brain dataset, MM-test accurately delineates fine-scale structures that are challenging for other methods, such as the 3D architecture of the pyramidal layer of the hippocampal cornu ammonis and the dentate gyrus. Theoretical guarantees—including selection consistency, FDR control, and an error bound for post-selection clustering—are also established.

\end{abstract}


\spacingset{1.9}
\section{Introduction}

The advent of spatial transcriptomics (ST) technologies has revolutionized gene expression analysis, enabling comprehensive transcriptome profiling while preserving crucial spatial information~\citep{marx2021method}. These cutting-edge techniques facilitate the simultaneous measurement of transcriptome-wide expression across thousands of tissue locations (``spots"), revealing previously inaccessible relationships between gene expression patterns and tissue architecture. By preserving spatial context, ST offers unprecedented insights into cellular organization and function, with its applications transforming developmental biology, neuroscience, and oncology research \citep{tang2026interpretable}. The primary analytical objective of ST technologies is the identification of spatial domains---tissue regions characterized by distinct cellular compositions, structural patterns, and functional roles. The delineation of these domains is achieved predominantly through the cluster analysis of high-dimensional ST data \citep{dries2021giotto, hu2021spagcn}. 
However, a substantial methodological challenge emerges due to the abundance of genes lacking spatial relevance. These spatially irrelevant genes introduce statistical noise that potentially obscures the underlying biological signals and compromises clustering efficiency. Consequently, the development of robust feature-selection methods that systematically identify genes integral to the spatial domain architecture is essential for accurate tissue characterization.

Ideally, a feature-selection method for ST data, also known as detection for spatially variable genes (SVGs), should be capable of handling the following statistical and computational challenges: 

\begin{itemize}
    \item The absence of a priori knowledge regarding tissue spatial domains necessitates the development of unsupervised approaches that identify SVGs based exclusively on their inherent spatial expression patterns.
    \item The ultra-high-dimensional nature of ST data presents a needle-in-a-haystack scenario, wherein only a small subset of genes exhibits biologically meaningful spatial variation. 
    \item The count-based, zero-inflated, and over-dispersed natures of ST data necessitate the use of specialized statistical modeling methodologies to differentiate true biological variation from technical artifacts \citep{sun2020statistical}. 
    \item The growing availability of three-dimensional (3D) ST datasets that capture whole tissue or organ structures \citep{schott2024open,chen2022spatiotemporal} requires integrative analysis frameworks capable of jointly identifying consistent SVGs across the 3D space.
    \item Effective false discovery rate (FDR) control is essential for reliable inference in high-dimensional contexts and for downstream functional analyses of SVGs, such as gene ontology or pathway analysis. 
\end{itemize}
Addressing these challenges requires the development of robust statistical methods that combine spatial pattern detection with rigorous feature selection principles.

Recently, SVG selection has attracted significant attention in the bioinformatics and computational biology literature \citep{ chen2025benchmarking}. These methods either incorporate multiple spatial correlation metrics for SVG identification \citep{sun2020statistical, zhu2021spark, yuan2024heartsvg}, or employ spatial generalized linear mixed-effects models to detect SVGs with non-zero random spatial effects \citep{Svensson2018SpatialDEIO, sun2020statistical, li2021bayesian}. However, most of these methods are designed for single-slice analyses and lack the ability to integrate information across multiple 3D tissue slices, which is increasingly important in modern ST datasets. In addition, these methods generally lack theoretical guarantees, such as selection consistency or FDR control. Moreover, they are not rotation-invariant, meaning that different rotations of spatial coordinates can lead to inconsistent results \citep{su2025rotation}.

In the statistics literature, SVG selection can be viewed as a feature screening problem for clustering analysis. The theoretical foundations of feature screening were first established by \citet{fan2008sure} in the context of linear regression, which stimulated extensive methodological developments in supervised settings \citep{liu2015selective}. However, the unsupervised counterpart for clustering problems remains relatively underexplored, despite its growing importance in high-dimensional applications, particularly in the detection of SVGs in ST data. Among the available methods, a prominent class evaluates features based on their distributional modality \citep{chan2010using,jin2016influential, banerjee2017feature,wang2025feature}, wherein unimodal distributions indicate cluster irrelevance and multimodal distributions suggest cluster relevance. More recently, \citet{liu2022clustering} introduced spectral clustering with feature selection (SCFC), a method that quantifies feature relevance through correlation with preliminary cluster assignments. These screening methods generally have well-established theoretical guarantees. However, they do not incorporate spatial or image-based context, which is essential for detecting spatial domains in ST data, and many do not provide any FDR control.

To address the challenges  of feature selection in ST data and facilitate domain detection, we introduce an MM-test statistic that assesses homogeneous mean models across clusters by leveraging the mean-variance relationship through a quasi-likelihood framework (Section~\ref{sec:setup0}). The test statistic was derived by maximizing the difference in quasi-likelihood between heterogeneous and homogeneous mean models via the majorization-minimization (MM) algorithm \citep{hunter2004tutorial}. The MM-test incorporates auxiliary information via a distance matrix that captures sample relationships using spatial or structural features. This matrix informs local mean estimation and dispersion modeling, enabling the MM-test to adapt to underlying spatial structures and improve sensitivity in identifying cluster-relevant features.
Feature selection is subsequently performed by retaining those variables whose MM-test statistics exceed a predefined threshold, $t_n$. 
We developed a knockoff filtering procedure that selects the threshold \(t_n\) to control the FDR.
The selected feature subset then enables cluster identification through conventional clustering approaches, such as k-means or spectral clustering algorithms. 

Next, we investigated the theoretical properties of the MM-test statistic and the feature-screening procedure in Section~\ref{sec:theory}. The screening procedure based on the MM-test achieved selection consistency, as established in Theorem~\ref{thm:consistency}. Theorem~\ref{coro:Hamm} shows that the Hamming error of the clustering results from our post-selection clustering analysis converged to zero asymptotically. In addition, the FDR of the MM-test was effectively controlled at a desired level, as demonstrated in Theorem~\ref{thm:FDR}.

Through comprehensive simulation studies, we demonstrate that the MM-test achieved superior accuracy in identifying cluster-relevant features across diverse scenarios in Section \ref{sec:simu}.  Subsequent clustering analysis of selected features exhibited significantly improved accuracy and robustness compared to baseline methods. In Section~\ref{sec:RealBench}, we benchmarked the MM-test against existing methods for detecting SVGs using 34 ST datasets. The results demonstrate that the MM-test consistently outperformed existing approaches. In Section~\ref{sec:3DApplication}, we applied MM-test to a 3D mouse brain datasets. We found that MM-test more accurately detected marker genes of known brain regions and hence enabled the identification of fine-scale 3D brain structures.

\section{Data description and motivation}\label{sec:data}
This study mainly analyzes a dataset derived from a whole-brain molecular atlas of the adult mouse brain, generated using spatial transcriptomics as described by \citet{ortiz2020molecular}. The dataset consists of 20 coronal brain slices spanning the anterior–posterior axis (Figure~\ref{fig:data}A; see Appendix K for details). Each slice was computationally registered to the Allen Mouse Brain Reference Atlas (\url{www.brain-map.org
}), yielding three-dimensional spatial coordinates for all measured tissue locations. The resulting three-dimensional dataset contains expression profiles for 11,879 genes across 11,061 spatially mapped tissue locations, forming an ultra–high-dimensional data structure. Besides, the sequenced gene expressions are discrete counts, rather than continuous numbers. At the same time, most genes exhibit zero or very low expression in any given location. On average, each tissue location has 4,510 detected genes and 14,739 total counts (Figure~\ref{fig:data}B). This reflects strong sparsity and small magnitude. Altogether, the combination of ultra–high dimensionality, sparsity, and small magnitude count data makes feature screening, particularly the identification of clustering-informative genes, a critical preprocessing step for downstream analysis.

Compared with conventional two-dimensional spatial transcriptomics experiments, this three-dimensional dataset provides an unprecedented view of gene expression patterns across the entire brain volume (Figure~\ref{fig:data}C). Many genes display spatially localized expression restricted to specific anatomical structures; such genes are commonly referred to as spatially variable genes (SVGs). Spatial SVGs carry rich information for delineating brain regions and subregions \citep{Zhang2023MolecularlyDA}, and some have been shown to reveal previously unrecognized fine-scale subdivisions \citep{qian2025spatial}. Because many anatomical regions extend broadly along the anterior–posterior axis, a single two-dimensional slice is insufficient to characterize SVGs in full spatial extent. For example, \textit{Ier5} shows strong spatially aggregation in cortex regions in three dimensional view (Figure~\ref{fig:data}C), but no such patterns in a two-dimensional slide (Figure~\ref{fig:data}D). Based on these three-dimensional SVGs, unbiased three-dimensional clustering enables reconstruction of complete region boundaries and corresponding region-specific gene expression patterns. Such integrated spatial-transcriptomic maps provide a foundation for constructing digital brain tissue models \citep{sharma2022digital}, which aim to jointly capture anatomical and molecular organization. These digital atlases have the potential to facilitate studies of brain function, disease mechanisms, and therapeutic target discovery \citep{shi2023spatial}.

\begin{figure}[htbp]
    \centering
    \includegraphics[width=1\linewidth]{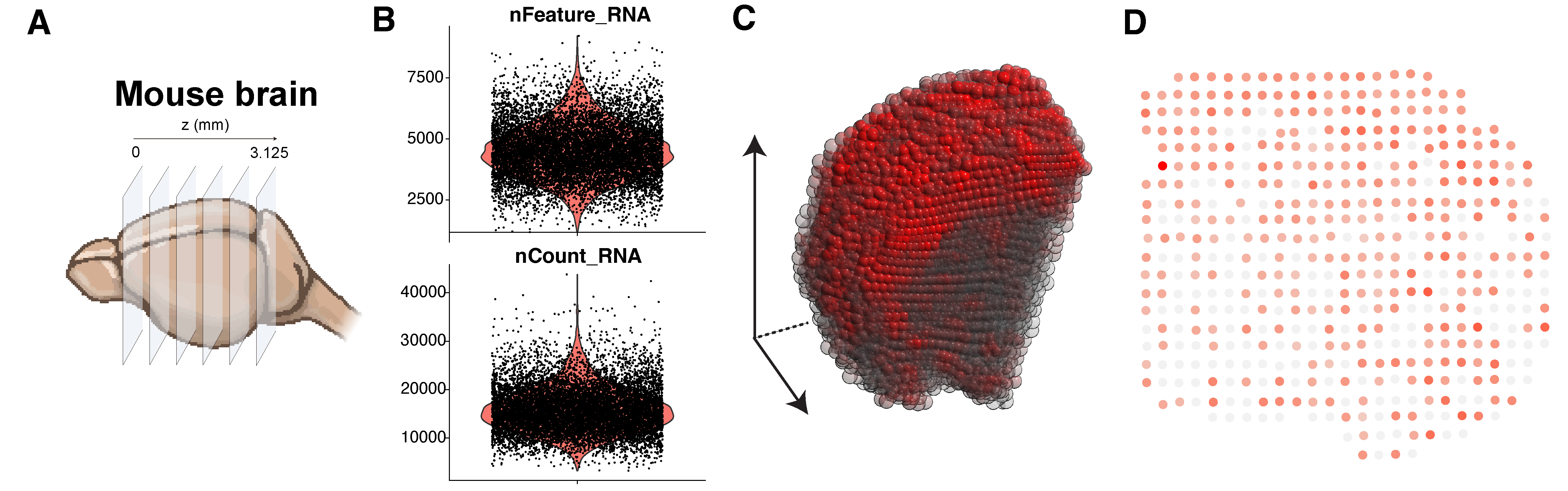}
    \caption{(A) Experimental design of the 3D spatial transcriptomics dataset from \citet{ortiz2020molecular}.
    (B) Violin plots of quality control metrics across samples.
The top panel shows nFeature\_RNA (detected genes per cell), and the bottom shows nCount\_RNA (total UMI counts). (C) 3D spatial visualization of \textit{Ier5} expression, which is enriched in the isocortex in this dataset. Color intensity indicates expression level, with red corresponding to higher expression. (D) Visualization of \textit{Ier5} expression in a two-dimensional slide.}
    \label{fig:data}
\end{figure}

\section{Model setup and the MM-test statistic}\label{sec:setup0}

\subsection{Model setup}\label{sec:setup}

Let \( [I] = \{1, \dots, I\} \) denote the index set associated with any integer \( I \). Consider \( n \) independent observations \( \mathbf{x}_i = (x_{i1}, \dots, x_{ip})^\top \in \mathbb{R}^p \), for each \( i \in [n] \), where each observation comes from one of \( K \) latent clusters.  
For a given cluster index \( k \in [K] \), define the set of sample indices belonging to cluster \( k \) as  
$
G_k = \{i \in [n] : k_i = k \},
$
where \( k_i \in [K] \) denotes the unknown cluster membership label for observation \( i \). Let  
$
\alpha_k = \frac{\# G_k}{n}
$
denote the empirical proportion of samples assigned to cluster \( k \), with \( \# G_k \) representing the cardinality of \( G_k \).

Let \( F_{kj} \) denote the distribution of the \( j \)th feature in the \( k \)th cluster. If observation \( i \) belongs to cluster \( k \), then the marginal distribution of \( x_{ij} \) is \( F_{kj} \), for all \( j \in [p] \) and \( i \in [n] \). Let \( \mu_{kj} \) and \( \tau_{kj} \) denote the mean and variance of \( F_{kj} \), respectively. That is, if \( X_{kj} \sim F_{kj} \), then \( \mu_{kj} = \mathbb{E}(X_{kj}) \) and \( \tau_{kj} = \operatorname{Var}(X_{kj}) \).
For a given feature \( j \), when the means \( \mu_{1j}, \dots, \mu_{Kj} \) are all equal across clusters, the feature does not contribute to distinguishing between clusters and is therefore considered \emph{cluster-irrelevant}. Thus, we define the sets of cluster-irrelevant and cluster-relevant features as follows:
\vspace{-10pt}
\begin{equation}\label{eq:set}
S_0 = \left\{ j : \mu_{1j} = \dots = \mu_{Kj} \right\}, \quad S_1 = \left\{ j : \mu_{kj} \ne \mu_{\ell j} \text{ for some } k, \ell \in [K] \right\}.
\vspace{-10pt}
\end{equation}
To identify the relevant features, we consider the following family of multiple hypothesis testing problems:
\vspace{-10pt}
\begin{equation}\label{eq:test}
\mathbb{H}_{j0} : \mu_{1j} = \dots = \mu_{Kj} \quad \text{vs.} \quad \mathbb{H}_{j1} : \mu_{kj} \ne \mu_{\ell j} \text{ for some } k, \ell.
\vspace{-10pt}
\end{equation}
We refer to the models under the null hypothesis \( \mathbb{H}_{j0} \) as \emph{homogeneous models}, and those under the alternative \( \mathbb{H}_{j1} \) as \emph{heterogeneous models}.
Since testing is performed feature-wise, for notational simplicity, we drop the subscript \( j \) and write \( F_k, \mu_k, \tau_k, \mathbb{H}_0, \mathbb{H}_1 \) instead of \( F_{kj}, \mu_{kj}, \tau_{kj}, \mathbb{H}_{j0}, \mathbb{H}_{j1} \), respectively.

A fundamental challenge in the testing problem \eqref{eq:test} that distinguishes it from classical hypothesis testing is that the cluster labels \( k_i \) are unobserved. One strategy to address this issue involves assuming that each \( F_k \) belongs to a known parametric family, thereby enabling likelihood-based tests to identify cluster-irrelevant features. However, these parametric methods can perform poorly if the true distributions deviate significantly from the assumed model.
As a flexible alternative, {distribution-free methods}, such as {quasi-likelihood approaches}, offer robust parameter estimation and valid inference without requiring the full specification of the data distribution.  
This makes them well-suited for complex or heterogeneous datasets, including spatial transcriptomics. 

A quasi-likelihood approach assumes the variance \( \tau_k \) is a known function of the mean \( \mu_k \), specifically \( \tau_k = V(\mu_k; \phi) \), where \( \phi > 0 \) is a dispersion parameter. The variance function \( V(\mu; \phi) \) typically takes the form  
$
\tau = V(\mu; \phi) = V_1(\mu) + \phi V_2(\mu),
$
where \( V_1(\mu) \geq 0 \) and \( V_2(\mu) > 0 \) are known continuously differentiable functions to the third order.
Common examples include the quasi-Poisson model (\( V_1(\mu) = 0 \) and \( V_2(\mu) = \mu \)) and the quasi-negative binomial model (\( V_1(\mu) = \mu \) and \( V_2(\mu) = \mu^2 \)). Notably, the variance function of the quasi-negative binomial model also corresponds to that of the Poisson mixture model, which is widely used in omics data analysis \citep{Sarkar2020SeparatingMA}.
The specified variance-mean function induces the distribution family $\mathcal{P}(\Theta, \Phi)$, defined as
\vspace{-10pt}
\begin{equation}\label{eq:calP}
	\mathcal{P}(\Theta, \Phi) = \left\{F:  X\sim F,\operatorname{Var}(X) = V_1(\Ex(X))+\phi V_2(\Ex(X)), \Ex(X) \in \Theta,  \phi \in \Phi  \right\},
    \vspace{-10pt}
\end{equation}
where \( \Theta \subset \mathbb{R} \) and \(\Phi \subset \mathbb{R}^+\) are compact parameter spaces. 
We then assume that each cluster-specific distribution \( F_k \) belongs to this family, i.e.,  \( F_k \in \mathcal{P}(\Theta, \Phi) \) for all \( k \in [K] \).

\subsection{MM-test statistic}

This section introduces the use of the MM-test statistic for evaluating the hypothesis framework (\ref{eq:test}). Conceptually, the MM-test statistic extends the likelihood ratio testing principle from parametric to semi-parametric inference by postulating only a functional relationship between the first two moments via a variance function $ V(\mu; \phi)$,  rather than specifying a full likelihood. 
Consequently, our test statistic was formulated by comparing quasi-likelihoods, instead of conventional likelihoods, between competing hypotheses. 
We began by establishing the quasi-likelihood framework \citep{wedderburn1974quasi}, which provides the theoretical basis for deriving the MM-test statistic. 

Let \( \mu = \mathbb{E}(x) \), \( \tau = \operatorname{Var}(x) \), and \( \phi > 0 \) denote the mean, variance, and dispersion, respectively.
We then define the \emph{quasi-likelihood} function as
\vspace{-10pt}
\[ Q(x; \mu, \phi) = \int_{x}^{\mu} \frac{x - t}{V(t; \phi)} \, \mathrm{d}t \vspace{-10pt}
 \]
and the \emph{quasi-density} function as 
\(f(x; \mu, \phi) = \exp(Q(x; \mu, \phi)). \)
Note that  \( f(x; \mu, \phi) \) is not a true probability density function, as its integral over the domain of \( x \) generally does not equal one. This quasi-likelihood construction, which preserves essential log-likelihood properties, facilitates inference even when the underlying probability model is incompletely specified.  

Let $f(x; \mu_k, \phi_k)$ denote the quasi-density function of the $k$th component ($k \in [K]$). We then define the overall \emph{quasi-density} function for the heterogeneous model as
\vspace{-10pt}
\[
\varphi(x;\bm{\alpha},\bm{\xi}, {\bm{\phi}})=\sum_{k=1}^{K}\alpha_k f(x; \mu_k, \phi_k),
\vspace{-10pt}
\]
where 
$\bm{\alpha}={(\alpha_1,\dots,\alpha_K)} , \bm{\xi}={(\mu_1,\dots,\mu_K)} \mbox{, and } \bm{\phi}={(\phi_1,\dots,\phi_K)}.$
Note that $\varphi(x;\bm{\alpha},\bm{\xi}, {\bm{\phi}})$ has a similar form to the density functions of standard mixture models. The \emph{quasi-likelihood} function for this heterogeneous model is subsequently defined as
\vspace{-10pt}
\[
 	 l_n(\bm{\alpha},\bm{\xi}, \bm{\phi})=\sum_{i=1}^{n}\log\ \varphi(x_i;\bm{\alpha},\bm{\xi},\bm{\phi}).
	 \vspace{-10pt}
\]
Under the homogeneous model $\mathbb{H}_0$, where $\mu_1 = \dots =  \mu_K = \mu_0$, the quasi-likelihood function becomes $l_n(\bm{\alpha},\bm{\xi}_0, \bm{\phi})$, with $\bxi_0 = (\mu_0, \dots, \mu_0)$. If all dispersion parameters $\phi_k$ are equal to a common $\phi$, we simplify the notation $l_n(\bm{\alpha},\bm{\xi}, \bm{\phi})$ to $l_n(\bm{\alpha},\bm{\xi}, \phi)$

Analogous to the classical likelihood ratio test, we define the quasi-likelihood ratio (QLR) test statistic as  
$
 \mathrm{QLR} = -2(L_0 - L_1),
 $  
where \( L_0 \) and \( L_1 \) are the maximum quasi-log-likelihoods under the null (\( \mathbb{H}_0 \)) and alternative (\( \mathbb{H}_1 \)) hypotheses, respectively. However, direct maximization of the quasi-log-likelihood is generally intractable due to the non-standard form of the quasi-likelihood function. To overcome this challenge, we employ the MM algorithm, a powerful optimization technique that iteratively refines parameter estimates by maximizing a surrogate function at each step (Appendix A.2). 
 
A challenge in applying the MM algorithm is that the update step for $\bm{\phi}$ lacks a closed form and is computationally intensive.  To address this, we simplify the optimization by constraining $\phi_1 = \dots = \phi_K = \phi$, and fixing $\phi$ as a working dispersion parameter $\hat{\phi}$ throughout the updates. 
The working dispersion parameter $\hat{\phi}$ is constructed by integrating auxiliary information such as spatial coordinates (Section \ref{subsec:overdispersion}). 
With this simplification, the MM updates become:
\vspace{-5pt}
 \begin{equation}\label{eq:updates}
  w_{ki}^{(t+1)}=\dfrac{\alpha_k^{(t)}{f(x_i; \mu_k^{(t)},\hat{\phi})}}{\varphi(x_i;\bm{\alpha}^{(t)},\bm{\xi}^{(t)}, \hat{\phi})},   \alpha_k^{(t+1)} = n^{-1}\sum_{i=1}^{n}w_{ki}^{(t+1)}, \mu_k^{(t+1)} = \frac{\sum_{i=1}^{n}w_{ki}^{(t+1)}x_i}{\sum_{i=1}^{n}w_{ki}^{(t+1)}}.
  \vspace{-5pt}
 \end{equation}
 Clearly, with the working dispersion $\hat{\phi}$,  the MM updates can be computed easily without additional optimization, making them computationally efficient. Furthermore, we demonstrate that the MM-test, defined based on the working dispersion $\hat{\phi}$, has strong theoretical guarantees for feature selection and downstream clustering (Section \ref{sec:theory}). 

After $T_n>0$ MM updates, we denote the maximized quasi-log-likelihood under \( \mathbb{H}_1 \) as $L_1=l_n\left(\bm{\alpha}^{(T_n)},\bm{\xi}^{(T_n)},  \hat{\phi}\right)$. 
Let $\hat{\bm{\xi}}_0=(\bar{\bx},\dots,\bar{\bx})$ be the estimator of $\bm{\xi}_0$, and let $l_n\left(\bm{{\alpha}}_0,\hat{\bm{\xi}}_0,  \hat{\phi}\right)$ represent the quasi-log-likelihood $L_0$ under \( \mathbb{H}_0 \). Then, the MM-test statistic is defined as
\vspace{-10pt}
$$
{\rm{MM}}_n^{(T_n)}=2\left\{l_n\left(\bm{\alpha}^{(T_n)},\bm{\xi}^{(T_n)},  \hat{\phi}\right)-l_n\left(\bm{{\alpha}}_0,\hat{\bm{\xi}}_0,  \hat{\phi}\right) \right\}.
\vspace{-10pt}
$$

Using the MM-test statistic, we propose the following procedure for cluster feature screening. Let ${\rm MM}_{nj}^{(T_n)}$ be the MM-test statistic for the $j$-th feature, corresponding to the hypothesis testing problem (\ref{eq:test}). Given the threshold $t_n>0$, we define the set of estimated cluster-relevant features, $\hat{S}_1(t_n) = \left\{  1\leq j \leq p: {\rm MM}_{nj}^{(T_n)} \geq t_n  \right\}$, which serves as an estimator for  $S_1$. A feature $j$ is included in $\hat{S}_1(t_n)$ if its MM-test statistic ${\rm MM}_{nj}^{(T_n)}\ge t_n$; otherwise, the feature is screened out. 
 


\subsection{Working dispersion $\hat{\phi}$ specification via spatial information} \label{subsec:overdispersion}
A key distinguishing factor between single-cell omics and spatial omics lies in the latter's incorporation of spatial context, typically represented through location graphs and often complemented by tissue images, which provide structural information beyond molecular expression profiles. This spatial information can be systematically leveraged to enhance cluster-relevant gene identification. We formalize the spatial information through an auxiliary distance matrix \( \mathbf{D} = [D_{ij}] \in \mathbb{R}^{n \times n} \), where \( D_{ii} = 0 \) for all \( i \), and \( D_{ij} > 0 \) for all \( i \neq j \). This matrix \( \mathbf{D} \) encodes the pairwise auxiliary distances between observations and is assumed to carry weak but informative signals about the underlying cluster structure, thereby enhancing feature selection beyond the ability of solely expression-based approaches. The construction of the \( \mathbf{D} \) matrix maintains flexibility and data-dependency, accommodating diverse information sources that capture clustering signals, including spatial coordinates (2D or 3D),
histological image features, and other modality-specific measurements. 

\begin{example}[Spatial transcriptomics data]\label{ex:spatial}
In spatial transcriptomics, spatial domains are identified by clustering 
the high-dimensional expression vectors $(x_{i1},\dots,x_{ip})$ at spatial spots, 
each with coordinates of either $(u_i,v_i)$ (2D) or $(u_i,v_i,w_i)$ (3D).  
A natural auxiliary distance between spots is the Euclidean distance, defined as:
$
D^S_{ij} = \sqrt{(u_i - u_j)^2 + (v_i - v_j)^2}
$
for two-dimensional spatial data, or
$
D^S_{ij} = \sqrt{(u_i - u_j)^2 + (v_i - v_j)^2 + (w_i - w_j)^2}
$
for three-dimensional spatial data. 

In addition, many spatial transcriptomics datasets include paired histology images, such as H\&E-stained images. We can use existing neural networks to extract a local image representation \( \bm{z}_i \in \mathbb{R}^d \) for each spot \( i \), and define an auxiliary distance \( D^I_{ij} \) as the Euclidean distance between \( \bm{z}_i \) and \( \bm{z}_j \):
$
D^I_{ij} = \left\| \bm{z}_i - \bm{z}_j \right\|_2.
$
Another option is to combine the spatial coordinates and histology images and define the distance as 
$D_{ij} = D^S_{ij} + D^I_{ij}.$
\end{example}

We leverage the auxiliary distance matrix  \( \mathbf{D} \) to obtain the working dispersion parameter \( \hat{\phi} \). The estimation procedure is preceded by its theoretical underpinnings. We define the pooled mean, variance, and dispersion as  
\vspace{-10pt}
\begin{equation}\label{eq:pooled}
\mu_0 = \sum_{k=1}^{K} \alpha_k \mu_k, \quad 
\tau_0 = \sum_{k=1}^{K} \alpha_k \tau_k + \sum_{k=1}^{K} \alpha_k (\mu_k - \mu_0)^2, \quad 
\phi_0 = \frac{\tau_0 - V_1(\mu_0)}{V_2(\mu_0)},
\vspace{-10pt}
\end{equation}
and let \( \bm{\xi}_0 = (\mu_0, \dots, \mu_0) \). Assuming that all clusters share the same dispersion parameter, i.e., \( \phi_1 = \cdots = \phi_K = \phi \), the quasi-Poisson model implies the quantity  
$
\phi = \phi_0 - \frac{ \sum_{k=1}^{K} \alpha_k (\mu_k - \mu_0)^2 }{ V_2(\mu_0) }
$, suggesting the dispersion parameter \( \phi \)  can be estimated as the difference between the pooled dispersion \( \phi_0 \) and 
the variance-scaled mean-difference sum of squares, where $\mu_k$ denotes cluster-specific means, $\mu_0$ the pooled mean, $\alpha_k$ the cluster proportions, and $V_2(\mu_0)$ the variance function. 

To approximate this relationship, we employ the auxiliary distance matrix \( \mathbf{D} \) to estimate the squared deviations between the local and global means. Specifically, for each sample \( i \), we define its neighborhood \( \mathfrak{N}(i) \) as the set of its \( r_n = n^{\beta} \) nearest neighbors, where \( 0 < \beta < 1 \) and \( \beta = 0.9 \) is used as the default value, determined by distance matrix \( \mathbf{D} \). The pooled dispersion ${\phi}_0$ can be estimated by 
$\hat{\phi}_0 = ({\hat{\tau}_0 - V_1(\bar{\mathbf{x}})})/{V_2(\bar{\mathbf{x}})},$ 
where $	\hat{\tau}_0=n^{-1}\sum_{i=1}^{n}(x_i-\bar{\mathbf{x}})^2$ is the sample variance. Thus, the dispersion parameter can be estimated by 
\vspace{-10pt}
\begin{equation}\label{eq:hatbetanonspa}
\hat{\phi} = \max\left\{ \hat{\phi}_0 - \sqrt{r_n} \cdot \frac{1}{n} \sum_{i=1}^{n} (\bar{\mathbf{x}} - \hat{\mu}_i)^2 \bigg/ V_2(\bar{\mathbf{x}}), \, m_\phi \right\},
\vspace{-10pt}
\end{equation}
where \( \hat{\mu}_i = \mathrm{mean}\{ x_\ell : \ell \in \mathfrak{N}(i) \} \), \( \sqrt{r_n} \) is a scaling factor that enhances detection power, and \( m_\phi > 0 \) is a constant that ensures the non-negativity of \( \hat{\phi} \).  

Intuitively, if the cluster information contained in the auxiliary data $\D$ is very strong, most of the neighborhoods $\frakN(i)$'s will consist only of samples from one cluster, and $\hat{\mu}_i$ will be close to one of $\mu_k$'s. Thus, $n^{-1}\sum_{i=1}^{n}  (\bar{\mathbf{x}}-\hat{\mu}_i)^2 $ is close to $\sum_{k=1}^{K} \alpha_k (\mu_0-\mu_{k})^2$. In general, the cluster information in $\D$ is weak, $n^{-1}\sum_{i=1}^{n}  (\bar{\mathbf{x}}-\hat{\mu}_i)^2 $ tends to significantly underestimate $\sum_{k=1}^{K} \alpha_k (\mu_{k} - \mu_0)^2$, so we add a scaling factor $\sqrt{r_n}$, which amplifies the signal under $\mathbb{H}_1$. 

Alternatively, under the null hypothesis \( \mathbb{H}_0 \), the feature is cluster-irrelevant: all clusters have the same mean, and the pooled dispersion \(\phi_0\) equals the dispersion \(\phi\). In this case, the local means \( \hat{\mu}_i \) also tend to be close to the pooled mean \(\mu_0\), and $\hat{\phi}_0$ is close to $\phi_0$. Therefore, even with the scaling factor $\sqrt{r_n}$, the empirical quantity \( \frac{\sqrt{r_n}}{n} \sum_{i=1}^{n} (\bar{\mathbf{x}} - \hat{\mu}_i)^2 \) is close to 0, and the working dispersion parameter \(\hat{\phi} \) is close to the dispersion $\hat{\phi}$. 

Therefore, under both null and alternative hypotheses, the working dispersion parameter \(\hat{\phi} \) serves as a reasonable surrogate for the true dispersion. Theoretical analysis shows that, with this working dispersion parameter, the MM-test achieves feature selection consistency and high clustering accuracy.

\subsection{FDR control via knockoff} \label{sec:Asym} 
A comprehensive feature-selection framework requires not only the identification of cluster-relevant features but also the assignment of valid $p$-values and, crucially, the control of FDR.
Recall that \( S_0 \subset [p] \) denotes the set of true null features (i.e., the cluster-irrelevant set), and \( \hat{S}_1 \) is the estimated set of cluster-relevant features. Let \( V = S_0 \cap \hat{S}_1 \) be the set of falsely rejected features. The FDR is formally defined as the expected proportion of false discoveries among all selections:
$
\mathrm{FDR} = \mathbb{E} \left( \frac{\# V}{\max \{ \# \hat{S}_1, 1 \}} \right).
$

Typically, the $p$-values of MM-test statistics are difficult to compute due to the non-convexity of the log-likelihood function.  To circumvent this inferential challenge, we developed the \emph{MM-test knockoff} procedure, a methodological extension of the $p$-value-free knockoff framework proposed by \citet{candes2018panning},  which enables rigorous FDR control without requiring explicit $p$-value computation or asymptotic approximations. 
Let \( \mathbf{X} = (\mathbf{X}_1, \dots, \mathbf{X}_p) = (\mathbf{x}_1, \dots, \mathbf{x}_n)^\top \in \mathbb{R}^{n \times p} \). The MM-test procedure involves the following steps:
\begin{enumerate}
    \item {\bf Knockoff Construction.} 
    For each feature \( j \in [p] \), we generate knockoff feature \( \widetilde{\mathbf{X}}_j \in \mathbb{R}^n \)  via sampling with replacement from \( \mathbf{X}_j \).  
    This yields the knockoff matrix \( \widetilde{\mathbf{X}} = (\widetilde{\mathbf{X}}_1, \dots, \widetilde{\mathbf{X}}_p) \in \mathbb{R}^{n \times p} \).
    
    \item {\bf MM-test Statistic Computation.} We calculate the MM-test statistic using the augmented matrix \( [\mathbf{X}, \widetilde{\mathbf{X}}] \in \mathbb{R}^{n \times 2p} \),  yielding 
    $
    \left( \mathrm{MM}_{n1}^{(T_n)}, \dots, \mathrm{MM}_{np}^{(T_n)}, \mathrm{MM}_{n1}^{(T_n)\star}, \dots, \mathrm{MM}_{np}^{(T_n)\star} \right),
    $
    and define scaled knockoff statistics for enhanced FDR control:
    \vspace{-10pt}
    \[
    \left( \widetilde{\mathrm{MM}}_{n1}^{(T_n)}, \dots, \widetilde{\mathrm{MM}}_{np}^{(T_n)} \right) = \left(1 + \frac{1}{\log n} \right) \left( \mathrm{MM}_{n1}^{(T_n)\star}, \dots, \mathrm{MM}_{np}^{(T_n)\star} \right).
    \vspace{-10pt}
    \]

    \item {\bf Threshold Determination.}
    For target FDR level \( q_0 \in (0, 1) \), we define the threshold \( H_{q_0} \):
    \vspace{-5pt}
    \begin{equation}\label{eq:Htresh}
    H_{q_0} = \min \left\{ h \in \mathcal{M}: \frac{ \#\{ j : \widetilde{\mathrm{MM}}_{nj}^{(T_n)} \geq h \} }{ \#\{ j : \mathrm{MM}_{nj}^{(T_n)} \geq h \}\vee 1 } \leq q_0 \right\},
    \vspace{-5pt}
    \end{equation}
    where \( \mathcal{M} = \{ |\mathrm{MM}_{nj}^{(T_n)}|: j \in [p] \} \) is the set of unique nonzero absolute test statistics and \( H_{q_0} = +\infty \) if no threshold satisfies the criterion.
\end{enumerate}

Our framework employs a simplified knockoff generation approach through direct resampling, circumventing the computational complexity of explicit joint distribution modeling \citep{candes2018panning}. As demonstrated in Section~\ref{sec:theory}, this computationally efficient procedure maintains theoretical guarantees for FDR control while enhancing practical implementation. The FDR-controlled feature selection set is formally defined using the threshold \( H_{q_0} \), 
$
\hat{S}_1(H_{q_0}) = \left\{ j \in [p] : \mathrm{MM}_{nj}^{(T_n)} \geq H_{q_0} \right\}.$
This data-adaptive threshold \( H_{q_0} \) provides empirically calibrated FDR control, making it particularly suitable for practical applications where the underlying feature distribution is unknown.

\section{Theoretical properties of the MM-test}\label{sec:theory}
Let \( \{a_n\} \) and \( \{b_n\} \) be positive sequences. We write \( a_n \ll b_n \) if \( a_n / b_n \to 0 \); \( a_n \gtrsim b_n \) if \( a_n \geq C|b_n| \) for some constant \( C > 0 \); similarly, \( a_n \lesssim b_n \) if \( a_n \leq C|b_n| \); and \( a_n \asymp b_n \) if both \( a_n \gtrsim b_n \) and \( a_n \lesssim b_n \) hold.

Before deriving the selection consistency guarantees for the MM-test feature screening procedure, we introduce four conditions (Conditions 1-4) that collectively ensure asymptotics. The complete mathematical formulations and detailed regularity conditions are provided in Appendix~B. 

Condition 1 is a moment constraint condition; it assumes that the \( L_q \) norm of each feature's distribution is uniformly bounded by the constant $M_q$, ensuring polynomial tail decay.
\footnote{If the distribution is assumed to have a sub-exponential tail, the probability bound in Theorem~\ref{thm:consistency} can be improved to exhibit an exponentially decaying rate.}
Condition 2 specifies a weak requirement on the spatial information and the between-cluster differences. Essentially, it assumes that, on average, the differences between the local means in the neighborhoods \( \mathfrak{N}(i) \) and the global mean \( \mu_0 \) exceed the threshold \( r_n^{-\vartheta} \), where \( r_n \) is the neighborhood size and \( \vartheta < 1/2 \) is a constant. This implies that the spatial data \( \mathbf{D} \) must contain a minimal level of cluster information; otherwise, most of the local means in \( \mathfrak{N}(i) \) will be very close to the global mean \( \mu_0 \). 

Condition 3 is a cluster separability condition that assumes the separation signal $\eta_n$ among the \( K \) clusters is sufficiently strong.
Condition 4 is a null homogeneity condition that, for cluster-irrelevant features, requires both the mean and variance to be identical across different clusters to ensure valid FDR control in the knockoff procedure.

We now establish the theoretical guarantees of our feature screening procedure, deriving rigorous asymptotic properties that ensure model selection consistency.


\begin{theorem}\label{thm:consistency}
	Assume that Conditions 1--2 hold.  Let $0 < \varpi < 1/2-\vartheta$ be a constant. 
When  $n$ is sufficiently large, for $r_n^{\vartheta}(\log n)^2 < T_n < r_n^{1/2-\varpi} \log n  $ and
$t_n = r_n^{-\kappa}, -1 + 3\vartheta <  \kappa < 5/2,$
we have 
$
\Pr\left( S_1 \neq  \hat{S}_1(t_n)\right) \lesssim (p-s) r_n^{-c_1q + \beta^{-1}} + sr_n^{-c_2q+\beta^{-1}} + o(1), 
$
where  $r_n = n^\beta, s = \# S_1$,  and $ c_1, c_2$ are two positive constants that only depend on $\kappa$ and $\vartheta$. 
\end{theorem}

Theorem \ref{thm:consistency} provides  the  probability of $S_1 \neq  \hat{S}_1(t_n)$, where $S_1$ represents the set of cluster-relevant features, and $\hat{S}_1(t_n)$ represents the features selected by our method in the spatial case.  
 If $p 
 \asymp n^{\omega}$ with $0<\omega + 1<q\beta(c_1\wedge c_2)$, we have $ \Pr\left( S_1 =  \hat{S}_1(t_n)\right)  \rightarrow 1,$ as $n \rightarrow \infty$, or in other words, we can achieve model selection consistency. 

We next derive the theoretical guarantees for clustering accuracy when applied to the MM-test-selected feature subset. Let \( \hat{\mathbf{k}}^{\mathrm{MM}} \) denote the vector for estimated cluster assignments obtained from k-means clustering in the PCA-reduced subspace of MM-test–selected features \citep{jin2016influential}.  
Parameter \( \eta_n \) quantifies the minimum between-cluster separation relative to within-cluster variation (formal definition provided in Appendix~A), serving as the clustering signal strength that governs classification accuracy. 
To quantify clustering accuracy, we employ the Hamming distance between the estimated and true cluster assignments,
  $
  \operatorname{Hamm}^*\left(\hat{\bf{k}}^{\mathrm{MM}}, \bf{k}\right)=\min _\pi\left\{\sum_{i=1}^n \pr\left(\hat{k}_{i}^{\mathrm{MM}} \neq \pi\left(k_i\right)\right)\right\},
  $
 where $\pi$ is any permutation in $\{1, \ldots, K\}$.  

  \begin{theorem}\label{coro:Hamm}
	Assume that $s = n^{\gamma}$ and cluster separation signal $\eta_n \gtrsim  s^{1/2}n^{-\vartheta}, 0\leq \vartheta < 1/4.$ For any $\gamma > 4\vartheta$ and $q > (\epsilon \gamma)^{-1} \vee (\gamma/\epsilon)$,  where $\epsilon = 1/4 - \max\{\vartheta, \vartheta/\gamma\} $, when  $n$ is sufficiently  large,  under Condition 1--3, we have: 
		$n^{-1}\operatorname{Hamm}^*\left(\hat{\bf{k}}^{\mathrm{MM}}, \bf{k}\right) \lesssim  \Pr\left\{ \hat{S}_1(t_n) \neq S_1 \right\} + o(1).$
\end{theorem}
Theorem~\ref{coro:Hamm} states that when clustering is performed using selected features, the Hamming distance between the resulting and true cluster assignments can be upper bounded by the probability of model mis-selection, plus a negligible error term introduced by the clustering procedure itself. This second term corresponds to the oracle clustering error obtained using all clustering-relevant features. Therefore, the clustering result based on the proposed method is nearly as accurate as that achieved using the full set of relevant features.
The derivation of Theorem~\ref{coro:Hamm} presents significant technical challenges, as our framework accommodates heavy-tailed distributions satisfying only polynomial moment conditions, and substantially generalizes previous results that rely on sub-Gaussian or sub-exponential assumptions prevalent in high-dimensional statistics \citep{jin2016influential}. The complete mathematical derivations with accompanying technical lemmas are presented in Appendix G, where we establish both finite-sample guarantees and asymptotic properties.
We now establish rigorous theoretical guarantees for FDR control under the MM-test procedure. 
\begin{theorem}\label{thm:FDR}
Assuming Conditions 1--4 hold, for any \( q_0 > 0 \), when $s \to \infty$, choosing the threshold \( H_{q_0} \) as defined in \eqref{eq:Htresh} ensures that
	$\mathrm{FDR} \leq  q_0 + o(1).$
\end{theorem}

Theorem~\ref{thm:FDR} states that the FDR is asymptotically controlled at the target level \( q_0 \) as the sample size \( n \to \infty \). The proof relies on establishing asymptotic distributional convergence between the original and knockoff statistics. 
Our proof strategy addresses the significant technical challenges that arise from the bootstrap-based knockoff generation, which only guarantees asymptotic, rather than exact, exchangeability between features and their knockoffs, and requires novel concentration inequalities and a discretization technique.   
The complete proof is provided in Appendix H.

\section{Simulation} \label{sec:simu}

In this section, we assess the performance of the MM-test by simulating spatial transcriptomics data. 
We conducted comparative analyses against a set of widely used methods \citep{chen2024evaluating}, including Moran \citep{miller2021characterizing}, SPARK-X \citep{zhu2021spark}, Binspect \citep{dries2021giotto} and SOMDE \citep{hao2021somde} for SVG detection and SCFS \citep{liu2022clustering} for feature screening. 
For the MM-test statistic, we used the default variance function \( V(\mu) = \mu + \phi \mu^2 \).
To assess the robustness of the MM-test to variance misspecification, we considered an alternative specification $V(\mu) = \phi\mu$. Complete simulation details of the robustness analysis and computational time comparisons across methods are provided in Appendix~J.1 and J.2, respectively. Appendix~J.3 introduces a three‐dimensional simulation designed to further examine the performance of the MM test statistic. 
In the simulation, we considered two distinct spatial layouts (Figure~S1).
Layout 1 is a square-shaped region consisting $n = 900$ spots, partitioned into $K = 5$ equal-sized rectangular domains.
Layout 2 is derived from annotated spatial transcriptomics data of a sagittal section of the mouse brain, consisting of $n = 2,687$ spots distributed across $K = 8$ anatomically-defined regions.
 In all simulations, we considered three dimensional setups, $p=3,000$, $10,000$ or $30,000$. 
Cluster-relevant/irrelevant features were generated from negative binomial (NB) distributions 
or Poisson log normal (PLN) distributions. 
 We randomly set 30\% of the elements to zero to increase the sparsity in the simulations. 
We considered two signal strength settings that quantify the differences between groups under the low and high signal strength conditions, respectively. 
 In each simulation setup, we generated 100 datasets. Detailed simulation settings are provided in Appendix I. Due to their high computation burden, Binspect and Moran were excluded from the evaluations of Layout 2 configurations with$p=10,000$ or $p=30,000$.

\subsection{Performance on feature screening}\label{subsec:FS}

We evaluated the feature screening performance through the area under the precision-recall curve (AUPRC) analysis. Features were ranked by their respective test statistics or $p$-values, and precision-recall curves were generated by varying these rank thresholds. 
The MM-test demonstrated superior or comparable performance in all simulation scenarios, with its advantages becoming particularly pronounced under more challenging conditions: complex spatial layouts (Layout 2), lower signal strengths, and higher feature dimensions (Table S1). Notably, in the most challenging scenario (Layout 2, low signal strength, $p=30,000$ features), the MM-test maintained robust performance, achieving a mean AUPRC of 0.75.
Next to the MM-test, SPARK-X exhibited the second-best performance (0.42), being reasonably effective across most scenarios. 

Next, Table \ref{tab:FDR} summarizes the statistical power (the ratio of correctly retained features to $s$) and the FDR for each method.
For the MM-test, features are retained if selected by the knockoff procedure with $q_0 = 0.05$.
For SCFS, cluster-relevant features are selected using its default parameters.  For SPARK-X, Binspect, SOMDE, and Moran, features are retained if their Benjamini–Hochberg-adjusted $p$-values are less than $0.05$. Compared with other methods, MM-test generally retained more true cluster-relevant features while incurring very few false positives. SPARK-X and SCFS tended to be conservative, exhibiting few false positives but often missing many true cluster-relevant features, especially in more challenging settings. The complexity of the spatial layout substantially influenced the performance of Moran and Binspect. These methods performed reasonably well in the simpler Layout 1 simulations but struggled with the more complex Layout 2 scenarios. 
The simulations further demonstrated that MM-test effectively controlled the FDR at the nominal level of 0.05, validating our statistical theory and offering a data-driven method for threshold determination.

\begin{table}[t]
	\centering
    \footnotesize
	\caption{The mean power and FDR of different methods over 100 replications. The target FDR level is 0.05. The numbers in the parentheses are the standard deviations of power and FDR  over 100 replications. The values in the table are the actual values $\times ~ 100$. Moran  and Binspect were not evaluated for $p=10,000$ and $p=30,000$ in Layout 2 simulations because they are computationally too expensive for these scenarios.}
			\setlength{\tabcolsep}{2mm}
	\begin{tabular}{cccccccc}
    \hline
    \hline
		$p$     &       & SCFS & SPARK-X & Moran & Binspect & SOMDE & MM-test \\
					\hline 
					\multicolumn{8}{c}{Layout-1, high signal} \\  
					\hline  
		\multirow{2}[0]{*}{3,000} & Power & 99 (4.5) & 100 (0.5) & 100 (0.1) & 96 (1.6) & 16 (1.9) & 100 (0.0) \\
		& FDR   & 0 (0.0) & 0 (0.0) & 30 (2.7) & 41 (3.2) & 83 (1.9) & 5 (2.3)  \\
		\multirow{2}[0]{*}{10,000} & Power & 98 (3.3) & 100 (0.6) & 100 (0.2) & 98 (1.0) & 15 (1.8) & 100 (0.0) \\
		& FDR   & 0 (0.0) & 0 (0.0) & 45 (2.3) & 61 (2.2) & 83 (1.9) & 5 (1.8) \\
		\multirow{2}[0]{*}{30,000} & Power & 69 (5.2) & 98 (1.1) & 100 (0.2) & 99 (0.8) & 6 (2.2) & 100 (0.0)  \\
		& FDR   & 0 (0.0) & 0 (0.0) & 62 (1.9) & 77 (1.1) & 92 (2.8) & 5 (2.1)  \\
							\hline 
		\multicolumn{8}{c}{Layout-1, low signal} \\  
		\hline  
		\multirow{2}[0]{*}{3,000} & Power & 0 (0.3) & 55 (3.3) & 49 (4.6) & 27 (3.9) & 0 (0.1) & 90 (3.4)  \\
		& FDR   & 0 (0.0) & 0 (0.0) & 38 (3.9) & 61 (4.3) & 10 (30.3) & 6 (2.8)\\
		\multirow{2}[0]{*}{10,000} & Power & 0 (0.0) & 42 (3.5) & 40 (4.2) & 30 (3.9) & 0 (0.0) & 69 (5.7)  \\
		& FDR   & 0 (0.0) & 0 (0.0) & 58 (3.0) & 79 (2.2) & 0 (0.0) & 6 (2.7) \\
		\multirow{2}[0]{*}{30,000} & Power & 0 (0.0) & 33 (2.9) & 35 (4.5) & 32 (3.4) & 0 (0.0) & 52 (6.9) \\
		& FDR   & 0 (0.0) & 0 (0.0) & 76 (3.0) & 90 (1.2) & 2 (14.1) & 5 (3.0)  \\
					\hline 
\multicolumn{8}{c}{Layout-2, high signal} \\  
\hline  
		\multirow{2}[0]{*}{3,000} & Power & 20 (0.7) & 69 (1.5) & 22 (1.6) & 6 (1.2) & 43 (1.5) & 100 (0.1) \\
		& FDR   & 0 (0.0) & 0 (0.0) & 0 (0.2) & 0 (0.0) & 9 (2.8) & 2 (1.0)  \\
		\multirow{2}[0]{*}{10,000} & Power & 15 (1.4) & 62 (1.6) & NA& NA & 8 (0.9) & 98 (0.7) \\
		& FDR   & 0 (0.0) & 0 (0.0) & NA & NA & 83 (1.9) & 3 (1.0)  \\
		\multirow{2}[0]{*}{30,000} & Power & 0 (0.0) & 54 (1.5) & NA & NA & 6 (0.9) & 90 (1.7)  \\
		& FDR   & 0 (0.0) & 0 (0.0) & NA & NA & 83 (2.3) & 4 (1.6)  \\
							\hline 
		\multicolumn{8}{c}{Layout-2, low signal} \\  
		\hline  
		\multirow{2}[0]{*}{3,000} & Power & 0 (0.1) & 36 (1.9) & 0 (0.0) & 0 (0.2) & 0 (0.1) & 87 (10.0) \\
		& FDR   & 0 (0.0) & 0 (0.0) & 0 (0.0) & 0 (0.0) & 0 (0.0) & 5 (2.2) \\
		\multirow{2}[0]{*}{10,000} & Power & 0 (0.0) & 28 (1.7) & NA & NA & 0 (0.0) & 66 (2.5)  \\
		& FDR   & 0 (0.0) & 0 (0.0) & NA & NA & 0 (0.0) & 6 (2.4)  \\
		\multirow{2}[0]{*}{30,000} & Power & 0 (0.0) & 23 (1.6) & NA & NA & 0 (0.0) & 57 (6.9)  \\
		& FDR   & 0 (0.0) & 0 (0.0) & NA & NA & 2 (14.1) & 4 (1.8)  \\
		\hline
	\end{tabular}%
	\label{tab:FDR}%
\end{table}%

\subsection{Feature screening improves clustering analysis}

After feature screening with FDR control, we further performed $k$-means clustering on the ten principal components derived from the reduced feature set. The number of clusters was fixed to match the true cluster number. Table \ref{tab:ARIspa} shows the adjusted Rand index (ARI) values between method-derived clusters and ground-truth groupings. 
MM-test consistently outperformed other methods across all scenarios,  particularly excelling in low-signal, high-dimensional, or spatially complex configurations. 
While the no-screening approach showed moderate success in lower dimensions, its performance deteriorated markedly in high-dimensional settings, underscoring the importance of rigorous feature screening before clustering. SPARK-X exhibited stable performance, but its accuracy was significantly lower than that of MM-test under demanding simulation conditions, reflecting its limitations in resolving subtle cluster boundaries.
SCFS underperformed in most scenarios primarily because its excessively conservative feature retention criteria disproportionately excluded true cluster-associated features.

\begin{table}[htbp]
	\footnotesize
	\centering
	\caption{The mean ARI of different methods over 100 replications. The numbers in the parentheses are the standard deviation of ARI over 100 replications.  The values in the table are the actual values $\times ~ 100$. Moran  and Binspect were not evaluated for $p=10,000$ and $p=30,000$ in Layout 2 simulations because they are computationally too expensive for these scenarios.}
			\setlength{\tabcolsep}{1mm}
	\begin{tabular}{ccccccccccc}
		\hline
        \hline
		$p$     & No-Screening & Oracle & SCFS  & SPARK-X & Moran & Binspect & SOMDE & MM-test \\
									\hline 
		\multicolumn{9}{c}{Layout-1, high signal} \\  
		\hline  
		3,000  & 95 (5.3) & 100 (0.2) & 99 (3.4) & \textbf{100 (0.2)} & \textbf{100 (0.3)} & \textbf{100 (0.4)} & 53 (4.0) & \textbf{100 (0.2)}  \\
		10,000 & 81 (4.2) & {100 (0.2)} & \textbf{100 (1.5)} & \textbf{100 (0.2) }& \textbf{100 (0.3)} & \textbf{100 (0.3)} & 51 (4.9) & \textbf{100 (0.2)}  \\
		30,000 & 53 (1.9) & 100 (0.3) & 96 (4.4) & \textbf{100 (0.3)} & \textbf{100 (0.3)} & 99 (0.4) & 23 (8.7) & \textbf{100 (0.3)} \\
											\hline 
		\multicolumn{9}{c}{Layout-1, low signal} \\  
		\hline  
		3,000  & 46 (2.1) & 85 (1.9) & 0 (0.8) & 62 (5.0) & 61 (4.2) & 36 (5.5) & 0 (0.0) & \textbf{84 (2.3) }\\
		10,000 & 4 (1.9) & 85 (2.4) & 0 (0.0) & 48 (5.0) & 51 (5.0) & 34 (5.1) & 0 (0.0) & \textbf{75 (3.7)}  \\
		30,000 & 1 (0.4) & 85 (2.1) & 0 (0.0) & 40 (3.6) & 41 (6.0) & 24 (6.4) & 0 (0.0) & \textbf{64 (6.5)} \\
									\hline 
\multicolumn{9}{c}{Layout-2, high signal} \\  
\hline  
		3,000  & 83 (9.0) & 97 (3.9) & 47 (8.5) & 85 (7.0) & 53 (6.4) & 31 (6.1) & 36 (5.0) & \textbf{98 (3.4)} \\
		10,000 & 49 (4.3) & 97 (4.1) & 41 (7.6) & 82 (6.2) & NA & NA & 33 (4.1) & \textbf{96 (5.0)}  \\
		30,000 & 4 (0.9) & 96 (5.1) & 0 (0.0) & 78 (6.7) & NA& NA& 11 (4.5) & \textbf{96 (5.3)}  \\
											\hline 
		\multicolumn{9}{c}{Layout-2, low signal} \\  
		\hline  
		3,000  & 42 (3.8) & 92 (1.6) & 0 (0.2) & 48 (7.0) & 0 (0.0) & 0 (0.7) & 7 (2.1) & \textbf{85 (8.3)}  \\
		10,000 & 6 (1.3) & 91 (3.2) & 0 (0.0) & 40 (6.6) & NA & NA & 7 (2.5) & \textbf{73 (6.1)}  \\
		30,000 & 0 (0.1) & 91 (3.6) & 0 (0.0) & 34 (6.7) & NA & NA & 2 (1.4) & \textbf{65 (8.0)}  \\
		\hline
	\end{tabular}%
	\label{tab:ARIspa}%
\end{table}%

\section{Benchmarking on real spatial transcriptomics data}\label{sec:RealBench}


In this section, we present benchmarking results for SVG identification methods on real spatial transcriptomics datasets, evaluating their performance in terms of detection accuracy, false positive control, and spatial clustering capability.
Following the benchmarking protocol established by \citet{chen2025benchmarking}, we assessed these methods across 34 publicly available spatial transcriptomics datasets with expert annotations of spatial domains, spanning diverse experimental protocols, tissue types, sample sizes, and spatial resolutions. The datasets include a human brain dataset (DLPFC) \citep{maynard2021transcriptome}, a developing mouse embryo dataset (DME) \citep{srivatsan2021embryo}, and an adult mouse brain (MAM) \citep{ortiz2020molecular}. 

To evaluate SVG detection accuracy, we adopted two ``silver standards'' of spatial domain marker genes, as considered by \citet{chen2025benchmarking}. One silver standard was generated by performing differential expression analysis between each annotated spatial domain and all other domains using the Wilcoxon rank-sum test. The other was constructed using NB regression of gene expression on spatial domains. In both approaches, genes with Benjamini–Hochberg adjusted p-values below 0.05 were selected as silver-standard genes.


\begin{figure}
    \centering
    \includegraphics[width=0.9\linewidth]{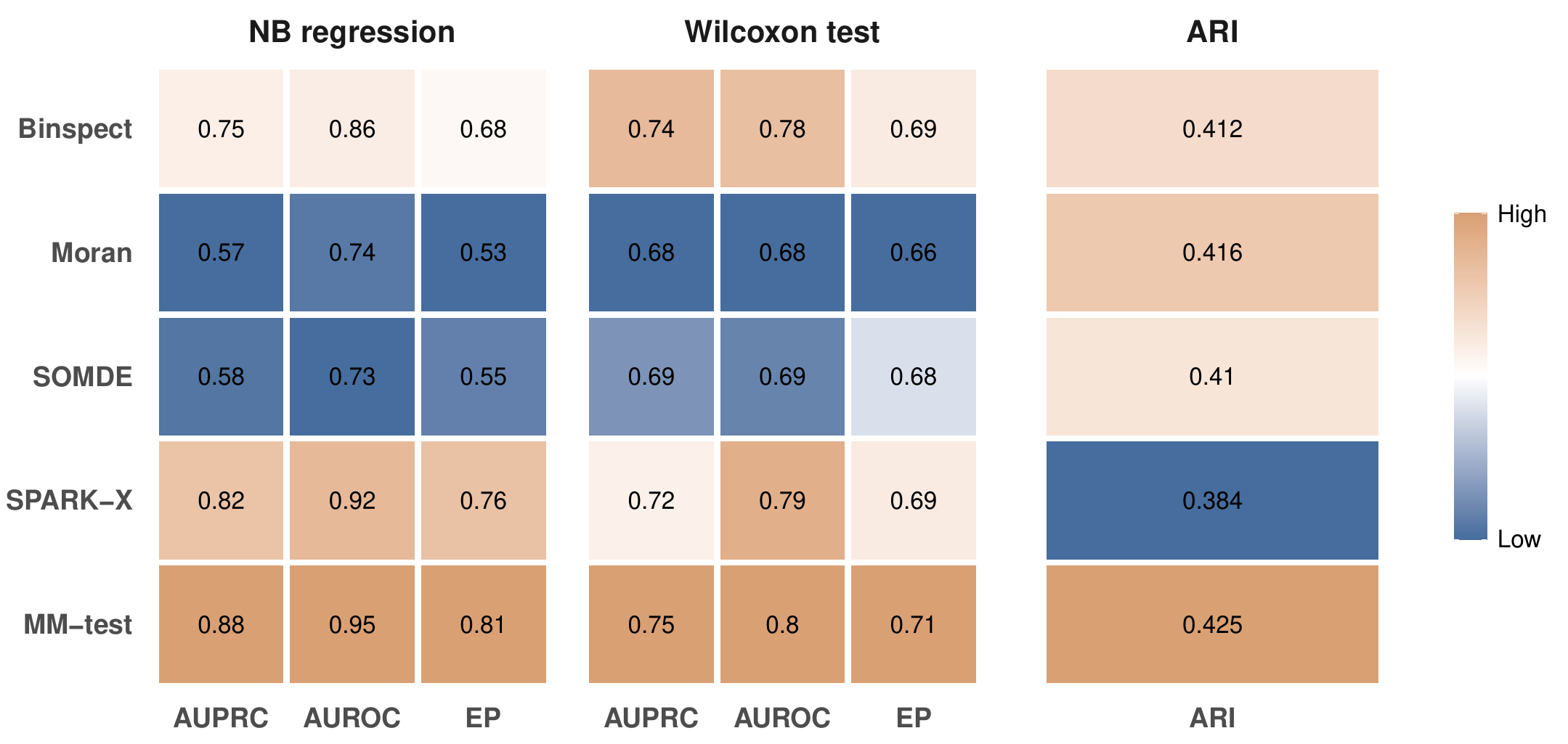}
    \caption{Benchmarking results of SVG identification methods across 34 real ST datasets. The y-axis represents the mean values of AUPRC, AUROC, and early precision (EP) for each method, calculated separately for each silver standard (Wilcoxon test, and negative binomial regression). The last panel shows the mean ARI values for spatial clustering based on the SVGs identified by each method. Higher values indicate better performance.}
    \label{fig:benchmark2}
\end{figure}

We adopted three evaluation metrics from \citet{chen2025benchmarking}: the area under the precision–recall curve (AUPRC), the area under the receiver operating characteristic curve (AUROC), and early precision (EP). Note that EP measures the fraction of true positives among the top-$K$ identified SVGs, where $K$ corresponds to the number of SVGs identified by each silver standard. Higher values in these metrics indicate better performance. 

Figure~\ref{fig:benchmark2} presents the mean AUPRC, AUROC, and EP values for each method across the 34 datasets, calculated separately for each silver standard. The MM-test consistently outperformed competing methods. It attained the highest AUPRC, AUROC, and EP scores under both the Wilcoxon test and NB regression criteria, underscoring its robustness in SVG detection. To evaluate false positive control in SVG detection, we conducted a null analysis using permuted datasets. As detailed in Appendix~K, MM-test, SPARK-X, and SOMDE effectively controlled false positives, whereas Moran and Binspect showed inflated false discovery rates, consistent with their performance in simulation studies (Figure~S3). 

Finally, we assessed the effectiveness of each method in facilitating spatial clustering based on the identified SVGs.  
Following the protocol of \citet{chen2025benchmarking}, we performed spatial clustering using the Seurat pipeline, optimized with the Louvain algorithm \citep{butler2018integrating}, based on the top 2,000 SVGs identified by each method. Clustering accuracy was quantified by the ARIs between the inferred clusters and expert-annotated spatial domains. To ensure fair comparison, the resolution parameter of the Louvain algorithm was varied from 0.5 to 1.5, and the value that maximized the ARI for each method was selected. As shown in Figure~\ref{fig:benchmark2}, the proposed MM-test achieved the highest ARI values, demonstrating its effectiveness in retaining spatially informative genes critical for domain identification.

\section{MM-test accurately detects fine-grained 3D spatial domains}\label{sec:3DApplication}

\begin{figure}[!htbp]
	\centering
	\includegraphics[width=0.9\linewidth]{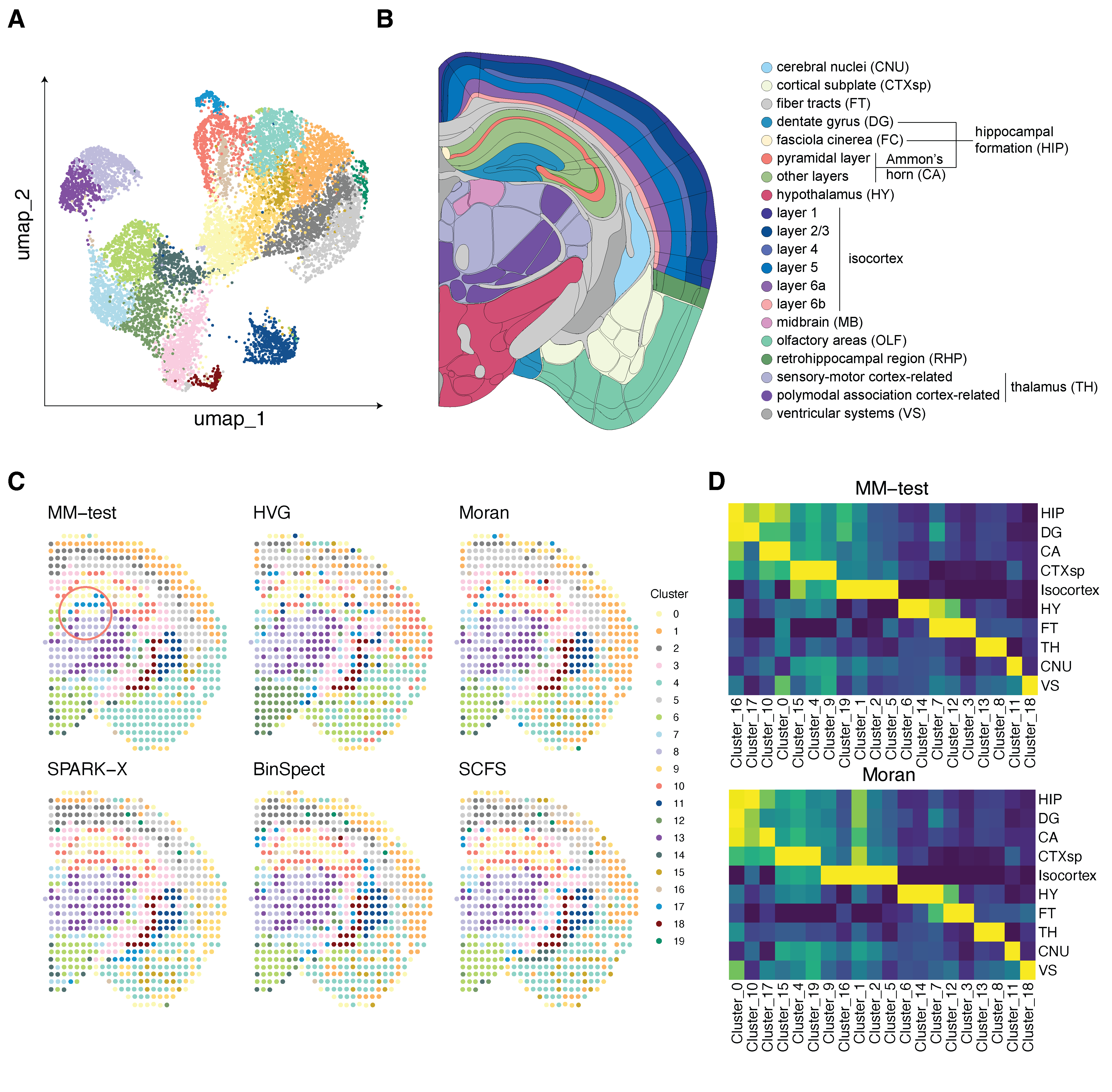}
\caption{ 
(A) UMAP visualization of cell clusters identified by the proposed MM-test method. 
(B) Brain region annotations based on the Allen Brain Atlas. 
(C) Clustering results on slice 23B from various methods, using their selected SVGs across all 20 slices. The red circle highlights the DG region, which was only clearly identified by MM-test. The arrow points to the DG region.
(D) Marker gene enrichment analysis: Heatmaps showing expression patterns of region-specific marker genes for clusters identified by MM-test and Moran. Marker genes are obtained from the Allen Brain Atlas \citep{lein2007genome}. Region abbreviations as defined in (B). 
}
	\label{fig:MM20}
\end{figure}

We applied MM-test to the 3D adult mouse brain dataset consisting of 20 coronal slices, as described in Section \ref{sec:data}. 
The auxiliary distance matrix of the MM-test was chosen as the 3D Euclidean distance between spatial locations. After gene selection using the knockoff procedure at the FDR level 0.05, we applied the Louvain algorithm to cluster the spatial locations with a resolution parameter 1.4. This procedure identified 20 spatial clusters (Figure~\ref{fig:MM20}A). For comparison, we adapted existing spatial gene detection methods—SPARK-X, Moran, and BinSpect—to operate in three dimensions by incorporating the 3D spatial coordinates of the spots. Additionally, we included two non-spatial feature selection methods: SCFS and the highly variable gene (HVG) approach \citep{butler2018integrating}. SOMDE was excluded from the comparison due to challenges in adapting it to 3D spatial data.  For all competing methods except HVG and SCFS, genes with Benjamini–Hochberg-adjusted \( p \)-values below 0.05 were selected. 
For SCFS, the default threshold of 0.9 was used to identify SVGs.
For HVG, the top 2,000 most variable genes were selected. The Louvain algorithm was also applied to each competing method, with resolution parameters adjusted such that 20 clusters were identified in each case (Figure~S4).

Overall, MM-test generated clusters that were more consistent with known brain domains (Figure \ref{fig:MM20}B,C). For example, the pyramidal layer of the hippocampal cornu ammonis (CAsp, Cluster 10) and the isocortex layers (Cluster 1, 2, 5 and 19) appeared clearer and more spatially continuous in the MM-test results compared to those from other methods. Using data from all 20 slices, all methods were able to distinguish the cerebral nuclei (CNU, Cluster 11) from neighboring domains such as the fiber tracts (FT, Cluster 3), ventricular systems (VS, Cluster 18), and the cortical subplate(CTXsp, Cluster 4). The thalamus (TH, Cluster 8 and 13) and hypothalamus (HY, Clusters 6 and 14) were also clearly delineated. Notably, two subregions of the thalamus were well separated: the sensory-motor cortex-related thalamus (DORsm, Cluster 8) and the polymodal association cortex-related thalamus (DORpm, Cluster 13). Marker gene enrichment analysis further revealed that MM-test clusters were enriched for region-specific marker genes, indicating strong concordance with known neuroanatomical annotations (Figure \ref{fig:MM20}D, S5). For instance, the isocortex, HY, TH and CNU clusters all showed high expression of their respective marker genes.

\begin{figure}[t]
	\centering
	\includegraphics[width=0.8\textwidth]{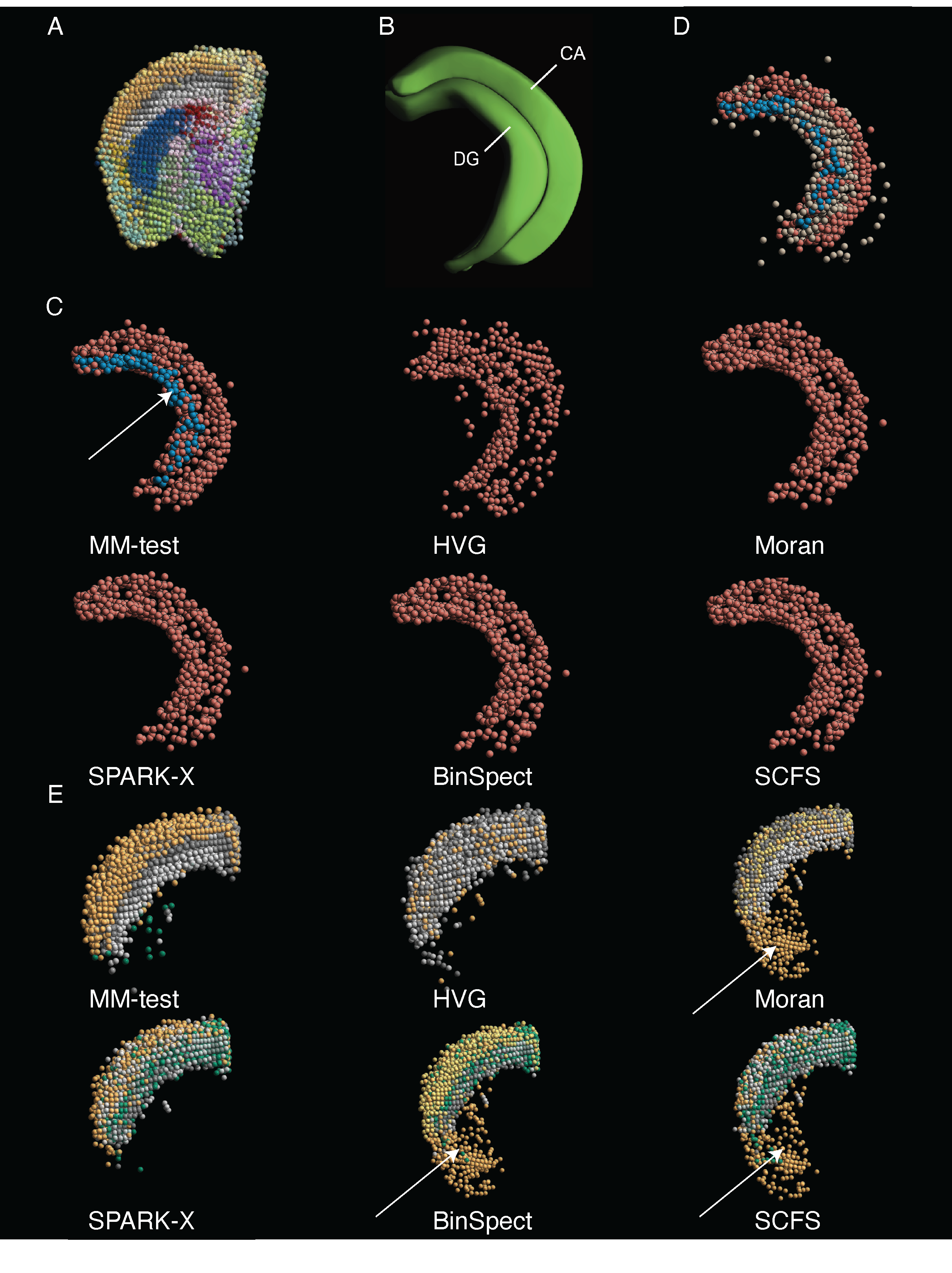}
\caption{
3D visualization of clusters identified by the MM-test across 20 consecutive slices. 
(A) All clusters identified by the MM-test. 
(B) 3D anatomical annotation of the CA and DG. 
(C) CAsp and DG regions (Cluster 10 and 17) identified by all methods. 
(D) CAsp (Cluster 10) and DG regions (Cluster 16 and 17) identified by all methods by MM-test.
(E) Isocortex region identified by all methods. The arrow points to the misclassified spots from CTXsp region (i.e. Moran's Cluster 1).
}
	\label{fig:3D}
\end{figure}

Interestingly, MM-test was the only method that clearly delineated the dentate gyrus (DG) cluster (cluster 17) and separated it from the CAsp cluster (Figure \ref{fig:MM20}C). Marker gene analysis confirmed that the DG cluster in MM-test highly expressed DG-specific marker genes, while the CAsp cluster expressed CA-specific markers (Figure \ref{fig:MM20}D, S6). In comparison, other methods failed to produce clusters with strong and specific expression of DG or CA marker genes. This distinction was particularly evident in the 3D visualization (Figure \ref{fig:3D}A), where MM-test revealed a much clearer boundary between the CAsp and DG clusters, aligning well with the 3D annotation from the Allen Brain Atlas (Figure \ref{fig:3D}B,C). The 3D visualizations from other methods further demonstrated their inability to distinguish the DG (Figure \ref{fig:3D}C). 

Notably, Cluster 16 also showed high expression of DG marker genes and may represent an additional DG cluster. While the 2D visualization (Figure~\ref{fig:MM20}C) showed only two points belonging to Cluster 16, the 3D visualization revealed that Cluster 16 and 17 are adjacent and together likely represent the DG (Figure~\ref{fig:3D}D). Additionally, the layered structure of isocortex clusters identified by MM-test appeared much clearer in 3D visualizations compared to those produced by other methods. Methods such as Moran, BinSpect, and SCFS misassigned many points within the isocortex clusters, incorrectly including many points likely belonging to CTXsp, indicating their reduced spatial specificity (Figure \ref{fig:3D}E).


We then conducted a sensitivity analysis by varying the number of slices included in the analysis. Specifically, we reduced the number of slices from 20 to 12, 5, and 1, and repeated the same analysis procedure. MM-test consistently identified refined clusters, such as the CAsp, DG, and CNU clusters (Table \ref{tab:spatial_comparison_multirow} and Figure S7--S9), when using between 20 and 5 slices of data. In comparison, the other methods failed to detect these clusters when using fewer than 20 slices. When using data from only a single slice, neither MM-test nor the competing methods were able to clearly delineate the DG and CNU clusters from their neighboring regions, highlighting the advantage of incorporating 3D multi-slice ST data. 

\begin{figure}[htbp]
	\centering
	\includegraphics[width=0.9\textwidth]{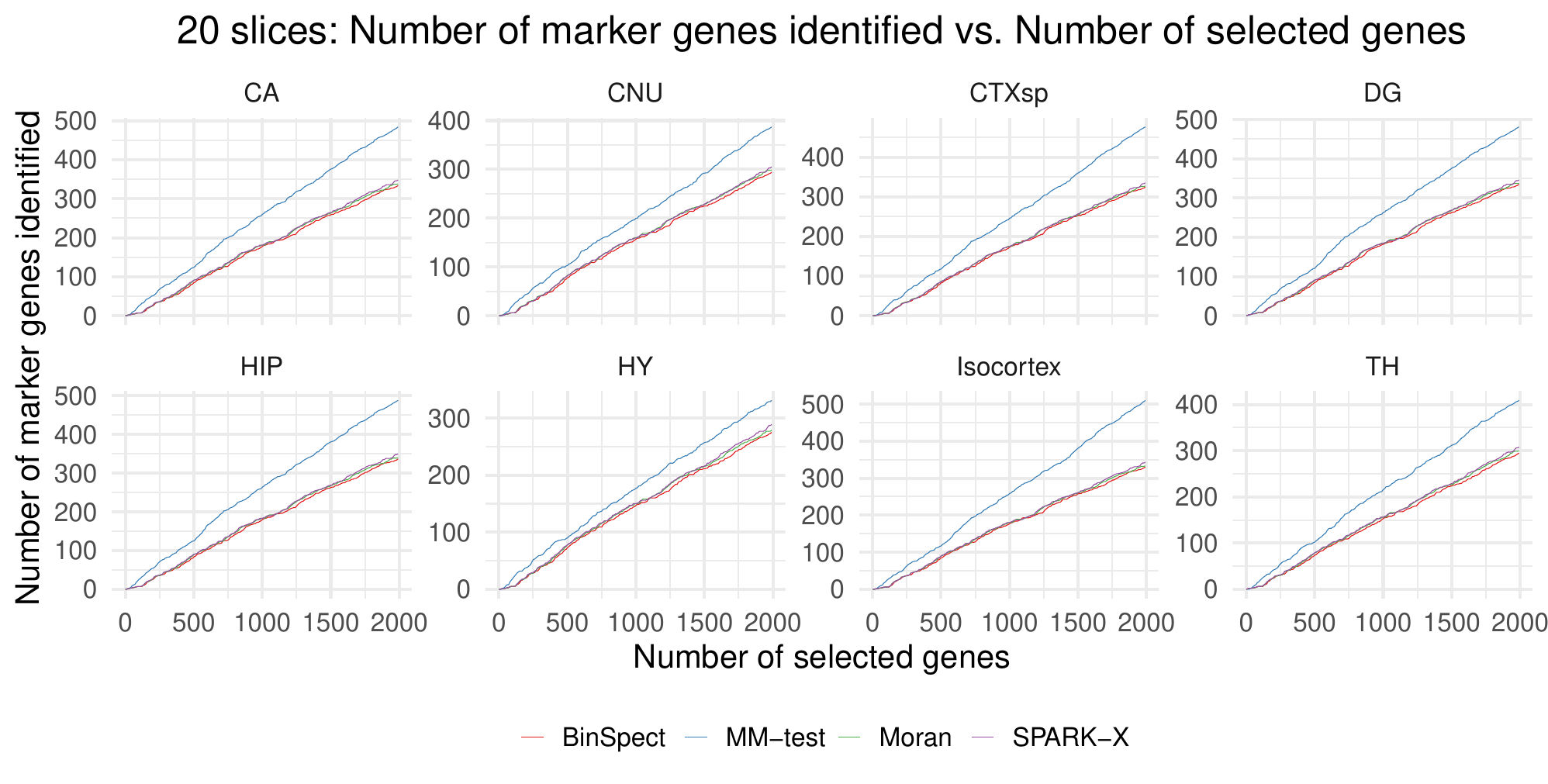}
\caption{Number of marker genes of different regions among top SVGs identified by different methods. Abbreviations as defined in Figure~\ref{fig:MM20}(B).
}
	\label{fig:bench20}
\end{figure}

Finally, to evaluate the accuracy of SVG selection by different methods, we compared top $K$ selected genes from each method with known marker genes from 12 brain regions curated from the Allen Brain Atlas \citep{lein2007genome}. Compared to the competing methods, MM-test consistently selected a greater number of known marker genes across different choice of $K$. For example, in the CAsp and DG regions, MM-test identified around 1.4 times more marker genes than other methods (Figure~\ref{fig:bench20}), contributing to improved clustering performance in these anatomically distinct domains. This advantage was also observed in the datasets with 12, 5, and 1 slices, although to a lesser extent (Figure S10--S12). These results suggest that the superior clustering performance of MM-test may be attributed to its higher SVG selection accuracy.

\section{Discussion}
SVG selection is a fundamental problem in ST data analysis. To address this, we develop MM-test, a distribution-free method that leverages auxiliary information to screen for SVGs. In spatial omics applications, we formulate the auxiliary information as a distance matrix that captures the physical proximity between tissue locations, allowing MM-test to be seamlessly applied to both 2D and 3D ST datasets. 

\begin{table}[htbp]
\centering
\caption{Sensitivity analysis with varying number of slices. The check mark (\cmark) indicates that the corresponding spatial domain was successfully identified by the given method, while a cross mark (\xmark) indicates that the domain was not detected.}
\footnotesize
\begin{tabular}{cccccccc}
\hline
\hline
{Number of slices} & {Spatial domain} & {MM-test} & {HVG} & {Binspect} & {Moran} & {SPARK-X} & {SCFS}  \\
\hline
\multirow{3}{*}{20} & CAsp and DG & \cmark & \xmark & \xmark & \xmark & \xmark & \xmark \\
                    & CNU & \cmark & \xmark & \cmark & \cmark & \cmark & \cmark \\
                  &  DORsm and 
                    DORpm & \cmark & \cmark & \cmark & \cmark & \cmark & \cmark \\
					\hline
\multirow{3}{*}{12} & CAsp and DG & \cmark & \xmark & \xmark & \xmark & \xmark & \xmark \\
                    & CNU & \cmark & \xmark  & \xmark  & \xmark  & \xmark  & \xmark \\
                                  &  DORsm and 
                    DORpm & \cmark & \cmark & \cmark & \cmark & \cmark & \cmark \\
					\hline
\multirow{3}{*}{5}  & CAsp and DG & \cmark & \xmark  & \xmark  & \xmark  & \xmark  & \xmark \\
                    & CNU & \cmark & \xmark  & \xmark  & \xmark  & \xmark  & \xmark \\
                                  &  DORsm and 
                    DORpm & \xmark & \xmark & \xmark & \xmark & \xmark & \xmark \\
\hline
\end{tabular}

\label{tab:spatial_comparison_multirow}
\end{table}
Unlike existing methods that primarily rely on spatial autocorrelation of gene expression, MM-test identifies SVGs by comparing expression means across unknown spatial clusters, thereby enabling robust detection of SVGs associated with complex spatial domain structures. Theoretically, we show that MM-test can distinguish features with heterogeneous means across clusters from those with homogeneous means with high probability, confirming its ability to select genes with different expressions across spatial domains. Furthermore, we establish an error bound for post-selection clustering, demonstrating that clustering based on MM-test-selected features achieves accuracy comparable to that obtained using the true clustering-relevant features. As a result, in real ST applications, MM-test more accurately detects marker genes associated with spatial domains and facilitates more precise spatial domain identification.

ST datasets are often accompanied by histology images, from which additional auxiliary information can be derived. By leveraging this information, MM-test can be extended beyond 3D multi-slice datasets to directly select SVGs in datasets containing multiple unrelated slices. In other applications, such as multi-omics integration, distance matrices can be constructed from complementary data modalities (e.g., proteomics, metabolomics, or epigenomics), allowing MM-test to be applied to feature selection beyond spatial transcriptomics. More broadly, MM-test offers a generalizable feature screening framework for applications where relational information among samples can be encoded as distances.

When the primary goal is to identify marker genes of spatial domains, MM-test combined with the knockoff procedure offers a statistically rigorous alternative to traditional two-step approaches that first perform spatial clustering and then differential expression analysis. Such conventional methods suffer from the ``double-dipping" problem, where using the same data for clustering and testing leads to invalid p-values and inflated FDR \citep{chen2024directly}. In contrast, MM-test directly selects spatially informative genes with rigorous FDR control, without relying on prior clustering. While the specific domains associated with the identified markers are initially unknown, they can be assigned post hoc to the clusters with the highest expression levels. This assignment step does not involve further testing and thus preserves the statistical validity of the marker selection.

\section{Data availability}
The R package \texttt{MMtestSVG}, along with the supporting code and datasets, is available on Zenodo at DOI: \href{https://doi.org/10.5281/zenodo.17070739}{10.5281/zenodo.17070739}.
\setlength{\bibsep}{0pt} 
\bibliographystyle{chicago}
\bibliography{reference}
 \end{document}